\documentclass{article}
\usepackage{amsmath}
\usepackage{amssymb}

\input{tcilatex}
\begin{document}

\author{J. G. Cardoso\thanks{%
jorge.cardoso@udesc.br} \\
Department of Mathematics\\
Centre for Technological Sciences-UDESC\\
Joinville 89223-100 SC\\
Brazil.
torsional spin-affine connexions; spin curvatures.}
\title{Two-Component Spinors in Spacetimes with Torsionful Affinities}
\date{}
\maketitle

\begin{abstract}
The essentially unique torsionful version of the classical two-component
spinor formalisms of Infeld and van der Waerden is presented. All the metric
spinors and connecting objects that arise here are formally the same as the
ones borne by the traditional formalisms. Any spin-affine connexion appears
to possess a torsional part which is conveniently chosen as a suitable
asymmetric contribution. Such a torsional affine contribution thus supplies
a gauge-invariant potential that can eventually be taken to carry an
observable character, and thereby effectively takes over the role of any
trivially realizable symmetric contribution. The overall curvature spinors
for any spin-affine connexion accordingly emerge from the irreducible
decomposition of a mixed world-spin object which in turn comes out of the
action on elementary spinors of a typical torsionful second-order covariant
derivative operator. Explicit curvature expansions are likewise exhibited
which fill in the gap related to their absence from the literature. It is
then pointed out that the utilization of the torsionful spinor framework may
afford locally some new physical descriptions.
\end{abstract}

\section{Introduction}

A remarkable property of Einstein-Cartan's gravitational theory [1-4] relies
upon the fact that the characteristic asymmetry of the Ricci tensor for any
torsionful world affine connexion always entails the presence of asymmetric
sources on the right-hand sides of the field equations. These sources were
traditionally identified [5, 6] with local densities of intrinsic angular
momentum of matter. Since the advent of Einstein-Cartan's theory, several
attempts have been made at designing torsional versions of extended
approaches to gravity that might circumvent the issues related to some
cosmological problems while supplying a macroscopic explanation of the
presently observable acceleration of the universe (see, for instance, Refs.
[7-10]). As brought forward by Refs. [11-15], torsional gravity has by
itself attracted much attention from researchers in conjunction with a
prediction accomplished by string theory that concerns the occurrence of
couplings between torsion and spinning fields. Amongst the developments that
have arisen from this field theoretical situation, noticeably enough, is the
work of Ref. [12] which provides a scheme that helps understand the ratios
between the coupling strengths for all the fundamental interactions. It is
thus shown that the value of a typical torsion-coupling constant can pass
through those of the coupling constants for the other interactions during
the cosmic evolution. Another noteworthy work posed in this connection [13],
allows for a family of leptons within a torsional gravitational framework
and establishes that the torsionic property of the underlying spacetime
geometry may give rise to interactions having the structure of the weak
forces. Moreover, for the case of a non-linear Lagrangian density for the
gravitational sector, the corresponding coupling configurations appear to
generate both the structure and the strength of the electroweak interactions
among leptons. It turned out, then, that the weak interactions among the
leptons effectively taken into consideration could be regarded as a
geometric effect due to couplings between torsion and spinor fields.

The two-component spinor framework for classical general relativity is
constituted by the so-called $\gamma \varepsilon $-formalisms of Infeld and
van der Waerden [16]. This framework was primarily aimed at describing the
dynamics of classical Dirac fields in curved spacetimes, with its
construction having been carried out much earlier than the achievement of
the definitive conditions for a curved space to admit spinor structures
locally [4]. Thus, the basic procedure just involves setting up two pairs of
conjugate spin spaces at every non-singular point of a curved spacetime that
is endowed with a torsionless covariant derivative operator. Furthermore,
the generalized Weyl gauge group [17] operates locally on any spin spaces in
a way that does not depend at all upon the action of the pertinent manifold
mapping group. One of the key assumptions lying behind the original
construction of the formalisms amounts to taking any Hermitian connecting
objects as covariantly constant entities. The implementation of this
assumption readily produces in either formalism a self-consistent set of
world-spin metric and affine correlations [18]. All the corresponding
curvature spinors arise most simply from the decomposition of mixed
world-spin quantities that result out of the action of covariant derivative
commutators on arbitrary spin vectors [19]. Loosely speaking, the most
striking physical feature of any such curvature spinors lies over the fact
that they are given as sums of purely gravitational and electromagnetic
contributions which bring forth in an inextricably geometric fashion the
occurrence of wave functions for gravitons and photons of both handednesses.
A fairly complete version of the $\gamma \varepsilon $-framework is given in
Ref. [18]. The gravitational contributions for the $\varepsilon $-formalism
were utilized in Ref. [20] to support a spinor translation of Einstein's
equations, but it had been established somewhat earlier [21] that any of
them should show up as a spinor pair which must be associated to the
irreducible decomposition of a Riemann tensor. In both the formalisms, any
gravitational wave functions turn out to be ultimately defined as totally
symmetric curvature pieces that occur in spinor decompositions of Weyl
tensors [20]. Any electromagnetic curvature contribution, on the other hand,
amounts to a pair of suitably contracted pieces\ that enter the spinor
representation of a locally defined Maxwell bivector and satisfy a peculiar
conjugation property [18, 19]. The propagation of gravitons for the $%
\varepsilon $-formalism, and the description of their couplings to external
electromagnetic fields, were given in Refs. [4, 20]. Nevertheless, only
recently [22, 23] has the full $\gamma \varepsilon $-description of the
propagation of spin curvatures in spacetime been obtained. It thus appears
that the couplings between gravitons and photons are exclusively borne by
the wave equations that govern the electromagnetic propagation. In Ref.
[24], it was likewise suggested that a description of some of the physical
properties of the cosmic microwave background may be achieved by looking at
the propagation in Friedmann-like conformally flat spacetimes of Infeld-van
der Waerden photons. The $\gamma \varepsilon $-framework was extensively
employed over the years by many authors in a more pragmatic way particularly
to reconstruct some classical generally relativistic structures and to
transcribe classification schemes for world curvature tensors [25-29].
However, the torsionful version of the formalisms has been sparsely
considered in the literature just to a minor extent [30, 31].

The present work exhibits systematically the natural torsional extension of
the classical $\gamma \varepsilon $-formalisms. One of the motivations for
elaborating our work is that it may certainly be of relevance for the
framework of modified gravity theories. We will assume at the outset that
the shift of any classical geometric considerations to the torsional context
must preserve both the structure of manifold mapping groups and the form of
the matrices that classically make out the Weyl gauge group. Hence, all the
defining prescriptions for the world and spin densities involved in the old
formalisms may be applicable equally well herein. Remarkably, the entire set
of algebraic configurations carrying the metric spinors and connecting
objects for the torsional formalisms, has the same form and gauge
characterizations as the one for the Infeld-van der Waerden formalisms. In
other words, the whole spinor algebra of the old framework is passed on
without any formal changes to the new framework. A typical spin-affine
connexion for either of the new formalisms carries additively a torsional
piece which is conveniently chosen as a suitable asymmetric contribution. In
contradistinction to any trivially realizable symmetric spin affinities,
such a torsional affine contribution thus supplies a gauge-invariant
potential that may carry an observable character. Because of the supposedly
legitimate additivity of spin affinities and the general pattern of
world-affine splittings, the classical world-spin affine correlations we had
referred to previously remain all formally valid within the torsional
framework, and thence also so do the classical covariant differential
expansions for world and spin densities as well as the system of Infeld-van
der Waerden metric eigenvalue equations [16]. The curvature spinors for some
spin-affine connexion occur in the decomposition of a characteristic mixed
world-spin object that accordingly comes from the action on elementary
spinors of a geometrically appropriate torsionful second-order covariant
derivative operator. Explicit curvature expansions are then obtained on the
basis of the use of well-known symbolic valence-reduction devices [4].
Indeed, the symmetry specification of the individual constituents of such
expansions were supplied in Ref. [31], but the overall curvatures for the
torsional framework have not been given hitherto.

Unless otherwise tacitly stated, the term "formalism" and the plural version
of it shall henceforward designate the new framework. The notation adopted
in Ref. [18] will be taken for granted except that spacetime components will
now be labelled by lower-case Greek letters. In particular, we denote as $%
x^{\mu }$ some local coordinates on a spacetime $\mathfrak{M}$ equipped with
a torsionful covariant derivative operator $\nabla _{\mu }$. The partial
derivative operator for $x^{\mu }$ is denoted as $\partial _{\mu }$. Any
world-metric tensors $g_{\mu \nu }$ and $g^{\mu \nu }$ on $\mathfrak{M}$
bear the traditional generally relativistic symmetry together with the local
signature $(+---)$, whence each of them still possesses $10$ real
independent components. We require $g_{\mu \nu }$ to fulfill the metric
compatibility condition%
\begin{equation*}
\nabla _{\mu }g_{\lambda \sigma }=0.
\end{equation*}%
Usually, $\mathfrak{M}$ should admit a local spinor structure in much the
same way as for the classical case of general relativity. Without any risk
of confusion, we will make use of the same indexed symbol $\nabla _{\mu }$
for expressing covariant derivatives in both formalisms. The elements of the
Weyl gauge group are non-singular complex $(2\times 2)$-matrices whose
components are defined as%
\begin{equation*}
\Lambda _{A}{}^{B}=\sqrt{\rho }\exp (i\theta )\delta _{A}{}^{B},
\end{equation*}%
where $\delta _{A}{}^{B}$ denotes the Kronecker symbol, $\rho $ stands for a
positive-definite differentiable real-valued function of $x^{\mu }$ and $%
\theta $ amounts to the gauge parameter of the group which is taken as an
arbitrary differentiable real-valued function on $\mathfrak{M}$. For the
determinant of $(\Lambda _{A}{}^{B})$, we have the expression 
\begin{equation*}
\det (\Lambda _{A}{}^{B})\doteqdot \Delta _{{\small \Lambda }}=\rho \exp
(2i\theta ).
\end{equation*}

It will be expedient to recall in Section 2 some facts concerning torsional
world geometry. This will considerably facilitate setting out some of the
spin properties of immediate interest to us. In spite of the fact that the
metric spinors and connecting objects for both formalisms are formally the
same as the ones for the traditional framework, we shall have to introduce
them into Section 3 along with the torsional spin affinities and some
world-spin affine configurations. There, the spin-metric algebraic
structures and affine devices for computing covariant derivatives of spin
densities are only slightly touched upon, but we will place emphasis on the
description of the patterns for the torsional affine contributions and their
behaviours under gauge transformations. The spin curvatures of the
formalisms are shown in Section 4. We draw an outlook from our work in
Section 5. A few additional conventions will be explained in due course.

\section{Torsional World Geometry}

The world affine connexion associated with $\nabla _{\mu }$ is split out as%
\begin{equation}
\Gamma _{\mu \nu \lambda }=\widetilde{\Gamma }_{\mu \nu \lambda }+T_{\mu \nu
\lambda },  \label{1}
\end{equation}%
where $\widetilde{\Gamma }_{\mu \nu \lambda }\doteqdot \Gamma _{(\mu \nu
)\lambda }$ may occasionally be identified with a Christoffel connexion, and 
$T_{\mu \nu \lambda }\doteqdot \Gamma _{\lbrack \mu \nu ]\lambda }$ is the
torsion tensor of $\nabla _{\mu }$. For some world-spin scalar $f$ on $%
\mathfrak{M}$, we have the differential prescription%
\begin{equation}
D_{\mu \nu }f=0,\text{ }D_{\mu \nu }\doteqdot 2(\nabla _{\lbrack \mu }\nabla
_{\nu ]}+T_{\mu \nu }{}^{\lambda }\nabla _{\lambda }),  \label{2}
\end{equation}%
whence the operator $D_{\mu \nu }$ is linear and possesses the Leibniz rule
property. It is obvious that $\widetilde{\Gamma }_{\mu \nu \lambda }$
carries $40$ real independent components whereas $T_{\mu \nu \lambda }{}$
carries $24$. The covariant derivatives of some purely world vectors $%
v^{\alpha }$ and $u_{\beta }$ are written down as 
\begin{equation}
\nabla _{\mu }v^{\lambda }=\text{ }\widetilde{\nabla }_{\mu }v^{\lambda
}+T_{\mu \sigma }{}^{\lambda }v^{\sigma },\text{ }\nabla _{\mu }u_{\lambda }=%
\text{ }\widetilde{\nabla }_{\mu }u_{\lambda }-T_{\mu \lambda }{}^{\sigma
}u_{\sigma },  \label{3}
\end{equation}%
where $\widetilde{\nabla }_{\mu }$ equals\footnote{%
The tensor $T_{\mu \lambda \sigma }{}$ of Ref. [4] conventionally equals $%
(-2)$ times ours.} the covariant derivative operator of $\widetilde{\Gamma }%
_{\mu \nu \lambda }$. For $v^{\lambda }$, for example, we have%
\begin{equation}
\widetilde{\nabla }_{\mu }v^{\lambda }\doteqdot \partial _{\mu }v^{\lambda }+%
\widetilde{\Gamma }_{\mu \sigma }{}^{\lambda }v^{\sigma }.  \label{3Lin}
\end{equation}%
When acting on world-spin scalars, the operators $\nabla _{\mu }$ and $%
\widetilde{\nabla }_{\mu }$ must agree with each other in the sense that
they should thus yield common results like $\partial _{\mu }f$.
Consequently, the metric compatibility condition for $g_{\mu \nu }$ can be
reexpressed as the expansion%
\begin{equation}
\widetilde{\nabla }_{\lambda }g_{\mu \nu }-2T_{\lambda (\mu \nu )}=0,
\label{4}
\end{equation}%
which is essentially equivalent to the relation%
\begin{equation}
\Gamma _{\mu }=\widetilde{\Gamma }_{\mu }+T_{\mu }=\partial _{\mu }\log (-%
\mathfrak{g})^{1/2},  \label{5}
\end{equation}%
with $\Gamma _{\mu }\doteqdot \Gamma _{\mu \lambda }{}^{\lambda }$, for
instance, and $\mathfrak{g}$ standing for the determinant of $g_{\mu \nu }$.

The Riemann tensor for $\Gamma _{\mu \nu \lambda }$ occurs in either of the
configurations%
\begin{equation}
D_{\mu \nu }v^{\lambda }=R_{\mu \nu \sigma }{}^{\lambda }v^{\sigma },\text{ }%
D_{\mu \nu }u_{\lambda }=-R_{\mu \nu \lambda }{}^{\sigma }u_{\sigma },
\label{6}
\end{equation}%
and obeys the equality%
\begin{equation}
R_{\mu \nu \lambda }{}^{\rho }=\widetilde{R}_{\mu \nu \lambda }{}^{\rho
}+R_{\mu \nu \lambda }^{{\footnotesize (T)}}{}^{\rho }+2(T_{[\mu \mid \tau
\mid }{}^{\rho }\widetilde{\Gamma }_{\nu ]\lambda }{}^{\tau }+\widetilde{%
\Gamma }_{[\mu \mid \tau \mid }{}^{\rho }T_{\nu ]\lambda }{}^{\tau }),
\label{7}
\end{equation}%
where the expressions for $\widetilde{R}_{\mu \nu \lambda }{}^{\rho }$ and $%
R_{\mu \nu \lambda }^{{\footnotesize (T)}}{}^{\rho }$ may be obtained from
the definition%
\begin{equation}
R_{\mu \nu \lambda }{}^{\rho }\doteqdot 2(\partial _{\lbrack \mu }\Gamma
_{\nu ]\lambda }{}^{\rho }+\Gamma _{\lbrack \mu \mid \tau \mid }{}^{\rho
}\Gamma _{\nu ]\lambda }{}^{\tau }),  \label{8}
\end{equation}%
just by putting the kernel letters $\widetilde{\Gamma }$ and $T$ in place of 
$\Gamma $, respectively. We should stress, however, that $R_{\mu \nu \lambda
}^{{\footnotesize (T)}}{}^{\rho }$ does \textit{not} constitute a tensor by
itself, but the sum of it with the crossed $\widetilde{\Gamma }T$-terms of
Eq. (\ref{7}) does. We will consider further this world characterization
later in Section 5.

It should be clear that $R_{\mu \nu \lambda \sigma }{}$ bears skewness in
the indices of each of the pairs $\mu \nu $ and $\lambda \sigma $, but the
Riemann-Christoffel index-pair symmetry does not take place here. Therefore,
\ $R_{\mu \nu \lambda \sigma }{}$ possesses $36$ real independent components
while its Ricci tensor possesses $16$. It can then be said that the Ricci
tensor for any affine connexion of the type specified by Eq. (\ref{1}),
carries asymmetry. Some symbolic computations easily show that the role of
the classical cyclic property of Riemann-Christoffel tensors has hereupon to
be taken over by%
\begin{equation}
R_{[\mu \nu \lambda ]}{}^{\sigma }-2\nabla _{\lbrack \mu }T_{\nu \lambda
]}{}^{\sigma }+4T_{[\mu \nu }{}^{\tau }T_{\lambda ]\tau }{}^{\sigma }=0,
\label{9}
\end{equation}%
whilst the Bianchi identity should now read%
\begin{equation}
\nabla _{\lbrack \mu }R_{\nu \lambda ]\sigma }{}^{\rho }-2T_{[\mu \nu
}{}^{\tau }R_{\lambda ]\tau \sigma }{}^{\rho }=0.  \label{10}
\end{equation}%
The property $R_{\mu \nu \lambda \sigma }=R_{[\mu \nu ][\lambda \sigma ]}$
and the torsionless relation $R_{[\mu \nu \lambda ]\sigma }{}=0$ entail
imparting\footnote{%
In Ref. [18], this is unclearly posed.} the index-pair symmetry to $R_{\mu
\nu \lambda \sigma }{}$. By invoking the dualization schemes given in Ref.
[4], and making some index manipulations thereafter, we rewrite Eqs. (\ref{9}%
) and (\ref{10}) as 
\begin{equation}
^{\ast }R^{\lambda }{}_{\mu \nu \lambda }+2\nabla ^{\lambda }{}^{\ast
}T_{\lambda \mu \nu }+4{}^{\ast }T_{\mu }{}^{\lambda \tau }{}T_{\lambda \tau
\nu }=0  \label{11}
\end{equation}%
and%
\begin{equation}
\nabla ^{\rho }{}^{\ast }R_{\rho \mu \lambda \sigma }+2{}^{\ast }T{}{}_{\mu
}{}^{\rho \tau }R_{\rho \tau \lambda \sigma }=0.  \label{12}
\end{equation}%
In general, such dualizations must take up the covariantly constant world 
\textit{tensors} $(-\mathfrak{g})^{1/2}\varepsilon _{\mu \nu \lambda \sigma
} $ and $(-\mathfrak{g})^{-1/2}\varepsilon ^{\mu \nu \lambda \sigma }$, with
these $\varepsilon $-objects being the Levi-Civitta world densities in $%
\mathfrak{M}$. Hence, according to Eq. (\ref{11}), the typical contracted
first-left dual pattern $^{\ast }R^{\lambda }{}_{\mu \lambda \nu }$ does not
vanish, in contrast to the Riemann-Christoffel case.

\section{Metric Spinors, Connecting Objects and Spin Affinities}

One of the fundamental metric spinors for the $\gamma $-formalism is
expressed by%
\begin{equation}
{\large (}\gamma _{AB}{\large )}=\left( 
\begin{array}{ll}
0 & \gamma \\ 
-\gamma & 0%
\end{array}%
\right) ,\text{ }\gamma =\mid \gamma \mid \exp (i\Phi ).  \label{13}
\end{equation}%
By definition, it behaves as a spin tensor under gauge transformations,
namely,%
\begin{equation}
\gamma _{AB}^{\prime }=\Lambda _{A}{}^{C}\Lambda _{B}{}^{D}\gamma
_{CD}=\Delta _{{\small \Lambda }}\gamma _{AB}.  \label{14}
\end{equation}%
The polar components $\mid \gamma \mid $and $\Phi $ are smooth real-valued
world scalars, with $\mid \gamma \mid \neq 0$ throughout $\mathfrak{M}$.
Their gauge behaviours will be described in a moment. For the inverse of $%
(\gamma _{AB})$, one finds the expression%
\begin{equation}
{\large (}\gamma ^{AB}{\large )}=\left( 
\begin{array}{ll}
0 & \gamma ^{-1} \\ 
-\gamma ^{-1} & 0%
\end{array}%
\right) ,  \label{15}
\end{equation}%
together with the component relationships%
\begin{equation}
\gamma _{AB}=\gamma \varepsilon _{AB},\text{ }\gamma ^{AB}=\gamma
^{-1}\varepsilon ^{AB},  \label{16}
\end{equation}%
where%
\begin{equation}
(\varepsilon _{AB})=\left( 
\begin{array}{ll}
0 & 1 \\ 
-1 & 0%
\end{array}%
\right) =(\varepsilon ^{AB}).  \label{17}
\end{equation}%
The $\varepsilon $-spinors of (\ref{17}) enter into the picture as
world-invariant entities subject to the laws%
\begin{equation}
\varepsilon _{AB}^{\prime }=(\Delta _{{\small \Lambda }})^{-1}\Lambda
_{A}{}^{C}\Lambda _{B}{}^{D}\varepsilon _{CD}=\varepsilon _{AB}  \label{18}
\end{equation}%
and 
\begin{equation}
\varepsilon ^{\prime AB}=\Delta _{{\small \Lambda }}\varepsilon ^{CD}\Lambda
_{C}^{-1}{}^{A}\Lambda _{D}^{-1}{}^{B}=\varepsilon ^{AB},  \label{19}
\end{equation}%
whence $\varepsilon _{AB}$ and $\varepsilon ^{AB}$ are invariant spin-tensor
densities of weights $-1$\ and $+1$, respectively. Thus, the entries $%
(\gamma ,\gamma ^{-1})$ appear as world-invariant spin-scalar densities of
weight $(+1,-1)$ and, consequently, both of $\gamma _{AB}$ and $\gamma ^{AB}$
bear world invariance as well. In accordance with these prescriptions, we
have the coupled laws%
\begin{equation}
\mid \gamma \mid ^{\prime }=\rho \mid \gamma \mid  \label{20}
\end{equation}%
and%
\begin{equation}
\exp (i\Phi ^{\prime })=\rho ^{-1}\Delta _{{\small \Lambda }}\exp (i\Phi ),
\label{21}
\end{equation}%
together with%
\begin{equation}
\partial _{\mu }^{\prime }\Phi ^{\prime }=\partial _{\mu }\Phi +2\partial
_{\mu }\theta .  \label{22}
\end{equation}

Any connecting objects for the $\gamma $-formalism satisfy anticommutation
relations of the form%
\begin{equation}
2\sigma _{AA^{\prime }(\mu }\sigma _{\nu )}^{BA^{\prime }}=\delta
_{A}{}^{B}g_{\mu \nu }.  \label{23}
\end{equation}%
For the $\varepsilon $-formalism, we have 
\begin{equation}
2\Sigma _{AA^{\prime }(\mu }\Sigma _{\nu )}^{BA^{\prime }}=\delta
_{A}{}^{B}g_{\mu \nu }.  \label{24}
\end{equation}%
The entries of the set\footnote{%
The kernel letter $S$ will henceforth stand for either $\sigma $ or $\Sigma $%
.} 
\begin{equation}
\mathbf{H}=\{S_{\mu AA^{\prime }},S_{AA^{\prime }}^{\mu },S_{\mu
}^{AA^{\prime }},S^{\mu AA^{\prime }}\},  \label{25}
\end{equation}%
are components of Hermitian $(2\times 2)$-matrices that depend smoothly upon 
$x^{\mu }$. Evidently, the Hermiticity of any element of the set (\ref{25})
is lost when we let its spinor indices share out both stairs. Some of the
most useful properties of the $S$-objects are expressed as%
\begin{equation}
S_{\mu A^{\prime }}^{(A}S_{\nu }^{B)A^{\prime }}=S_{A^{\prime }[\mu
}^{(A}S_{\nu ]}^{B)A^{\prime }}=S_{A^{\prime }[\mu }^{A}S_{\nu
]}^{BA^{\prime }},  \label{26}
\end{equation}%
whence we can likewise write%
\begin{equation}
S_{AA^{\prime }\mu }S_{\nu }^{AA^{\prime }}=S_{AA^{\prime }(\mu }S_{\nu
)}^{AA^{\prime }}.  \label{27}
\end{equation}%
The basic world-spin metric relations are thus given by%
\begin{equation}
g_{\mu \nu }{}=S_{\mu }^{AA^{\prime }}S_{\nu }^{BB^{\prime
}}M_{AB}M_{A^{\prime }B^{\prime }}  \label{28}
\end{equation}%
and 
\begin{equation}
M_{AB}M_{A^{\prime }B^{\prime }}=S_{AA^{\prime }}^{\mu }S_{BB^{\prime
}}^{\nu }g_{\mu \nu }{},  \label{29}
\end{equation}%
whereas the spinor structure that represents the tensor $(-\mathfrak{g}%
)^{1/2}\varepsilon _{\mu \nu \lambda \sigma }$ is written as\footnote{%
The symbol "c.c." has been taken throughout what follows to denote an
overall complex conjugate piece.}%
\begin{equation}
e_{AA^{\prime }BB^{\prime }CC^{\prime }DD^{\prime
}}=i(M_{AC}M_{BD}M_{A^{\prime }D^{\prime }}M_{B^{\prime }C^{\prime }}-\text{%
c.c.}),  \label{30}
\end{equation}%
with the kernel letter $M$ denoting here as elsewhere either $\gamma $ or $%
\varepsilon $. Every connecting object behaves as a vector as regards the
action of the manifold mapping group of $\mathfrak{M}$. Those for the $%
\gamma $-formalism bear a gauge-tensor character while the Hermitian ones
for the $\varepsilon $-formalism have to be regarded as invariant
spin-tensor densities carrying the absolute weights $\pm 1$. For instance,%
\begin{equation}
\sigma _{\mu A^{\prime }}^{\prime B}=\Lambda _{A^{\prime }}{}^{B^{\prime
}}\sigma _{\mu B^{\prime }}^{C}\Lambda _{C}^{-1}{}^{B}=\exp (-2i\theta
)\sigma _{\mu A^{\prime }}^{B}  \label{31}
\end{equation}%
and%
\begin{equation}
\Sigma _{AA^{\prime }}^{\prime \mu }=\rho ^{-1}\Lambda _{A}{}^{B}\Lambda
_{A^{\prime }}{}^{B^{\prime }}\Sigma _{BB^{\prime }}^{\mu }=\Sigma
_{AA^{\prime }}^{\mu }.  \label{32}
\end{equation}

We suppose that spin affinities in $\mathfrak{M}$ bear an additivity
property in both formalisms apart from the eventual implementation of any
symmetry splittings. Typically, in either formalism, we have%
\begin{equation}
\vartheta _{\mu AB}\doteqdot \widetilde{\vartheta }_{\mu AB}+\vartheta _{\mu
AB}^{{\footnotesize (T)}}.  \label{33}
\end{equation}%
In Eq. (\ref{33}), $\widetilde{\vartheta }_{\mu AB}$ is identified with the
spin-affine connexion for the torsionless operator $\widetilde{\nabla }_{\mu
}$ whilst $\vartheta _{\mu AB}^{{\footnotesize (T)}}$ accounts for the
torsionfulness of $\nabla _{\mu }$. Thus, for some spin vectors $\zeta ^{A}$
and $\xi _{A}$, we have the corresponding patterns%
\begin{equation}
\nabla _{\mu }\zeta ^{A}=\widetilde{\nabla }_{\mu }\zeta ^{A}+\vartheta
_{\mu B}^{{\footnotesize (T)}}{}^{A}\zeta ^{B},\text{ }\nabla _{\mu }\xi
_{A}=\widetilde{\nabla }_{\mu }\xi _{A}-\vartheta _{\mu A}^{{\footnotesize %
(T)}}{}^{B}\xi _{B}.  \label{34}
\end{equation}%
Towards making it feasible to ensure the self-consistency of the world-spin
metric and affine structures carried by $\mathfrak{M}$, it is seemingly
necessary to allow for the gauge-invariant constancy requirement 
\begin{equation}
\nabla _{\lambda }S_{AA^{\prime }}^{\mu }=0.  \label{35}
\end{equation}%
In both formalisms, the spacetime metric compatibility condition then gets
translated into%
\begin{equation}
\nabla _{\mu }(M_{AB}M_{A^{\prime }B^{\prime }})=0.  \label{36}
\end{equation}

The piece $\widetilde{\vartheta }_{\mu AB}$ of the prescription (\ref{33})
is required to carry only complex entries whence it contributes in either
formalism $32$ real independent components to $\vartheta _{\mu AB}$. It is
apparently suggestive to think of the torsional piece of Eq. (\ref{33}) as
having the symmetry property $\vartheta _{\mu AB}^{{\footnotesize (T)}%
}{}=\vartheta _{\mu (AB)}^{{\footnotesize (T)}}{}$. This choice would at
once supply the required $24$ real independent components if it were
actually taken into account. It has been used by some authors [4] for
carrying out a rough spinor transcription of Einstein-Cartan's theory. Even
though symmetry properties are gauge invariant, the symmetric choice for $%
\vartheta _{\mu AB}^{{\footnotesize (T)}}{}$ would nevertheless appear to be
inadequate insofar as Eqs. (\ref{35}) and (\ref{36}) remain both unaltered
when we add to each of $\widetilde{\vartheta }_{\mu A}{}^{B}$ and $\vartheta
_{\mu A}^{{\footnotesize (T)}}{}^{B}$ purely imaginary world-covariant
quantities\footnote{%
Such quantities should be the same in both formalisms. This was established
in Ref. [18] for the case of the classical framework.} of the type $\pm
i\iota _{\mu }\delta _{A}{}^{B}$. As far as the situation at issue is
concerned, the main point is that the implementation of a symmetric $%
\vartheta _{\mu AB}^{{\footnotesize (T)}}{}$ rules out all the possibilities
of sorting out contracted torsional affinities to which one could eventually
ascribe a physical meaning. Rather than implementing the symmetric torsional
choice, we should make use of an asymmetric prescription that provides $20$
real independent components along with four more coming from the purely
imaginary trace%
\begin{equation}
\vartheta _{\mu A}^{{\footnotesize (T)}}{}^{A}=-2iA_{\mu },  \label{37}
\end{equation}%
with $A_{\mu }$ thus being a world vector. A possible choice for $\vartheta
_{\mu A}^{{\footnotesize (T)}}{}^{B}$ is prescribed as 
\begin{equation}
(\vartheta _{\mu A}^{{\footnotesize (T)}}{}^{B})=\left( 
\begin{array}{cc}
a_{\mu } & b_{\mu } \\ 
\beta _{\mu } & c_{\mu }%
\end{array}%
\right) ,  \label{38}
\end{equation}%
with the conditions%
\begin{equation}
\func{Im}b_{\mu }=\func{Im}\beta _{\mu }=0,\text{ }\vartheta _{\mu A}^{%
{\footnotesize (T)}}{}^{A}=(a_{\mu }+c_{\mu })\doteqdot -2iA_{\mu }.
\label{39}
\end{equation}%
Here, we disregard any torsional spin-affine prescription having Re$%
\vartheta _{\mu A}^{{\footnotesize (T)}}{}^{A}\neq 0$, but every choice for $%
\vartheta _{\mu A}^{{\footnotesize (T)}}{}^{B}$ is gauge invariant (see Eq. (%
\ref{48}) below).

Of course, the patterns of $\vartheta _{\mu AB}^{{\footnotesize (T)}}{}$ and 
$\vartheta _{\mu A}^{{\footnotesize (T)}}{}^{A}$ as stipulated by Eqs. (\ref%
{37})-(\ref{39}) adequately carry $(20+4)$ real independent components in
all. For $\widetilde{\vartheta }_{\mu A}{}^{A}$, we similarly write the
world-covariant prescription%
\begin{equation}
\func{Im}\widetilde{\vartheta }_{\mu A}{}^{A}=-2\Phi _{\mu }.  \label{40}
\end{equation}%
Hence, in both formalisms, we have the common piece%
\begin{equation}
\func{Im}\vartheta _{\mu A}{}^{A}=-2(\Phi _{\mu }+A_{\mu }).  \label{41}
\end{equation}%
In the $\gamma $-formalism, Eq. (\ref{36}) right away yields the four-real
parameter relation%
\begin{equation}
\func{Re}\widetilde{\gamma }_{\mu A}{}^{A}=\partial _{\mu }\log \mid \gamma
\mid .  \label{42}
\end{equation}%
We observe that the right-hand side of (\ref{42}) bears world covariance as $%
\mid \gamma \mid $ is a world-invariant real spin-scalar density. In the $%
\varepsilon $-formalism, there occurs no metric relation like (\ref{42})
such that the respective piece $\func{Re}\widetilde{\vartheta }_{\mu
A}{}^{A} $ must be defined by hand such as in the classical framework (for
further details, see Ref. [18]). In effect, we have the world-covariant $%
\varepsilon $-contribution%
\begin{equation}
\func{Re}\widetilde{\Gamma }_{\mu A}{}^{A}=\Upsilon _{\mu }.  \label{43}
\end{equation}%
It follows that $\widetilde{\vartheta }_{\mu AB}{}$ and its trace contribute 
$(32+8)$ real independent components to the overall $\vartheta _{\mu AB}$.
The pieces $(\widetilde{\vartheta }_{\mu AB},\widetilde{\vartheta }_{\mu
A}{}^{A})$ and $(\vartheta _{\mu AB}^{{\footnotesize (T)}},\vartheta _{\mu
A}^{{\footnotesize (T)}}{}^{A})$ are thus quantities that carry $(32,8)$ and 
$(20,4)$ real independent components in either formalism, and therefore
recover the numbers of independent components of $\widetilde{\Gamma }_{\mu
\nu \lambda }$ and $T_{\mu \nu \lambda }$ appropriately.

To establish the gauge behaviours of any spin affinities for either
formalism, we implement the covariant property%
\begin{equation}
\nabla _{\mu }^{\prime }\xi _{A}^{\prime }=\Lambda _{A}{}^{B}\nabla _{\mu
}\xi _{B}.  \label{44}
\end{equation}%
Writing out the expansions of (\ref{44}) explicitly, after some differential
manipulations, we end up with the law%
\begin{equation}
\vartheta _{\mu A}^{\prime }{}^{B}=\vartheta _{\mu A}{}^{B}+\frac{1}{2}%
(\partial _{\mu }\log \Delta _{{\small \Lambda }})\delta _{A}{}^{B},
\label{45}
\end{equation}%
whence, making a contraction over the indices $A$ and $B$ carried by (\ref%
{45}), gives rise to%
\begin{equation}
\vartheta _{\mu A}^{\prime }{}^{A}=\vartheta _{\mu A}{}^{A}+\partial _{\mu
}\log \Delta _{{\small \Lambda }}.  \label{46}
\end{equation}%
By this point, we call for the old procedure whereby any two-component
spin-affine configurations should be built up so as to formally look like
world ones.\footnote{%
This procedure really underlies the construction of the classical framework.}
In both formalisms, the torsional piece $\vartheta _{\mu AB}^{{\footnotesize %
(T)}}$ should therefore behave covariantly under the action of the gauge
group, which means that%
\begin{equation}
\gamma _{\mu AB}^{{\footnotesize (T)}\prime }{}=\Delta _{{\small \Lambda }%
}\gamma _{\mu AB}^{{\footnotesize (T)}}{},\text{ }\Gamma _{\mu AB}^{%
{\footnotesize (T)}\prime }{}=(\Delta _{{\small \Lambda }})^{-1}\Lambda
_{A}{}^{C}\Lambda _{B}{}^{D}\Gamma _{\mu CD}^{{\footnotesize (T)}}{}=\Gamma
_{\mu AB}^{{\footnotesize (T)}}{}.  \label{47}
\end{equation}%
Thus, $\vartheta _{\mu A}^{{\footnotesize (T)}}{}^{B}$ must bear gauge
invariance in each formalism, that is to say,%
\begin{equation}
\vartheta _{\mu A}^{{\footnotesize (T)}\prime }{}^{B}=\vartheta _{\mu A}^{%
{\footnotesize (T)}}{}^{B}.  \label{48}
\end{equation}%
So, the behaviour of the torsionless piece $\widetilde{\vartheta }_{\mu
A}{}^{B}$ is such that it must absorb the inhomogeneous term lying on the
right-hand side of Eq. (\ref{45}), whence we should effectively combine (\ref%
{48}) with%
\begin{equation}
\widetilde{\vartheta }_{\mu A}^{\prime }{}^{B}=\widetilde{\vartheta }_{\mu
A}{}^{B}+\frac{1}{2}(\partial _{\mu }\log \Delta _{{\small \Lambda }})\delta
_{A}{}^{B}.  \label{49}
\end{equation}%
Calling upon Eqs. (\ref{14}) and (\ref{18}) then yields the laws%
\begin{equation}
\widetilde{\gamma }_{\mu AB}^{\prime }{}=\Lambda _{A}{}^{C}\Lambda _{B}{}^{D}%
\widetilde{\gamma }_{\mu CD}{}+\frac{1}{2}(\partial _{\mu }\Delta _{{\small %
\Lambda }})\gamma _{AB}  \label{50}
\end{equation}%
and%
\begin{equation}
\widetilde{\Gamma }_{\mu AB}^{\prime }{}=(\Delta _{{\small \Lambda }%
})^{-1}\Lambda _{A}{}^{C}\Lambda _{B}{}^{D}\widetilde{\Gamma }_{\mu CD}{}+%
\frac{1}{2}(\partial _{\mu }\log \Delta _{{\small \Lambda }})\varepsilon
_{AB}{}.  \label{51}
\end{equation}%
Obviously, the contracted versions of Eqs. (\ref{48}) and (\ref{49}) amount
to the same thing as%
\begin{equation}
A_{\mu }^{\prime }=A_{\mu },\text{ }\Phi _{\mu }^{\prime }=\Phi _{\mu
}-\partial _{\mu }\theta  \label{52}
\end{equation}%
and%
\begin{equation}
\func{Re}\widetilde{\vartheta }_{\mu A}^{\prime }{}^{A}=\func{Re}\widetilde{%
\vartheta }_{\mu A}{}^{A}+\partial _{\mu }\log \rho .  \label{53}
\end{equation}

The construction of the affine devices for computing covariant derivatives
of spin densities is based upon the requirement which amounts to looking
upon the $\varepsilon $-metric spinors as covariantly constant objects in
either formalism. This requirement is also implemented within the
traditional Infeld-van der Waerden framework. We thus allow for a
world-covariant quantity $\vartheta _{\mu }$ defined by%
\begin{equation}
\nabla _{\mu }\varepsilon _{AB}=0\Leftrightarrow \vartheta _{\mu }\equiv 
\widetilde{\vartheta }_{\mu }+\vartheta _{\mu }^{{\footnotesize (T)}%
}=\vartheta _{\mu A}{}^{A},  \label{54}
\end{equation}%
with $\widetilde{\vartheta }_{\mu }\doteqdot \widetilde{\vartheta }_{\mu
A}{}^{A}$ and $\vartheta _{\mu }^{{\footnotesize (T)}}\doteqdot \vartheta
_{\mu A}^{{\footnotesize (T)}}{}^{A}$. Therefore, it also occurs in the
formal configuration 
\begin{equation}
\nabla _{\mu }\gamma _{AB}=\nabla _{\mu }(\gamma \varepsilon
_{AB})=\varepsilon _{AB}\nabla _{\mu }\gamma ,  \label{55}
\end{equation}%
and likewise is made up by the expansion 
\begin{equation}
\nabla _{\mu }\gamma =\widetilde{\nabla }_{\mu }\gamma -\gamma \vartheta
_{\mu }^{{\footnotesize (T)}},  \label{56}
\end{equation}%
which constitutes the prototype in both formalisms for covariant derivatives
of complex spin-scalar densities of weight $+1$. Clearly, the right-hand
side of (\ref{56}) stands for a covariant expansion for the independent
component of $\gamma _{AB}$. For a complex spin-scalar density $\alpha $ of
weight $\mathfrak{w}$ in $\mathfrak{M}$, we then have%
\begin{equation}
\nabla _{\mu }\alpha =\widetilde{\nabla }_{\mu }\alpha -\mathfrak{w}\alpha
\vartheta _{\mu }^{{\footnotesize (T)}}.  \label{57}
\end{equation}%
In case a density $\beta $ carries the absolute weight $2\mathfrak{a}$, we
will get the real expansion%
\begin{equation}
\nabla _{\mu }\beta =\partial _{\mu }\beta -2\mathfrak{a}\beta \func{Re}%
\vartheta _{\mu }=\widetilde{\nabla }_{\mu }\beta ,  \label{58}
\end{equation}%
whence, from Eqs. (\ref{20}) and (\ref{42}), we see that $\mid \gamma \mid $
is covariantly constant in the $\gamma $-formalism. Hence, all the $\Sigma $%
-objects bear covariant constancy in both formalisms.

In fact, the recovery in either formalism of covariant derivative patterns
for arbitrary world tensors may only be achieved if the covariant constancy
property (\ref{35}) is accounted for. This condition allows us to deal with
combined world-spin displacements in $\mathfrak{M}$. For instance,%
\begin{equation}
\nabla _{\mu }u^{\lambda }=S_{AA^{\prime }}^{\lambda }\nabla _{\mu
}u^{AA^{\prime }}\Leftrightarrow \nabla _{\mu }u^{AA^{\prime }}=S_{\lambda
}^{AA^{\prime }}\nabla _{\mu }u^{\lambda },  \label{59}
\end{equation}%
where $u^{\lambda }$ amounts to a world vector. Some manipulations involving
rearrangements of the index configurations of (\ref{59}) then yield the
general $\gamma $-formalism relationship%
\begin{equation}
\Gamma _{\mu AA^{\prime }BB^{\prime }}+\sigma _{\lambda BB^{\prime
}}\partial _{\mu }\sigma _{AA^{\prime }}^{\lambda }=\gamma _{\mu AB}\gamma
_{A^{\prime }B^{\prime }}+\text{c.c.}.  \label{60}
\end{equation}%
Hence, by recalling Eq. (\ref{5}) and the spin-affine prescriptions given
before, we get the correlation%
\begin{equation}
4\func{Re}\widetilde{\gamma }_{\mu A}{}^{A}=\widetilde{\Gamma }_{\mu
}+T_{\mu }+\sigma _{\lambda }^{AA^{\prime }}\partial _{\mu }\sigma
_{AA^{\prime }}^{\lambda }.  \label{61}
\end{equation}%
In the $\varepsilon $-formalism, $u^{AA^{\prime }}$ is an Hermitian
spin-tensor density of absolute weight $+1$, and one has the expansion%
\begin{equation}
\nabla _{\mu }\Sigma _{AA^{\prime }}^{\lambda }=\partial _{\mu }\Sigma
_{AA^{\prime }}^{\lambda }+\Gamma _{\mu \nu }{}^{\lambda }\Sigma
_{AA^{\prime }}^{\nu }-(\Gamma _{\mu A}{}^{B}\Sigma _{BA^{\prime }}^{\lambda
}+\text{c.c.})+\Upsilon _{\mu }\Sigma _{AA^{\prime }}^{\lambda },  \label{62}
\end{equation}%
which can evidently be reset as%
\begin{equation}
\nabla _{\mu }\Sigma _{AA^{\prime }}^{\lambda }=\widetilde{\nabla }_{\mu
}\Sigma _{AA^{\prime }}^{\lambda }+T_{\mu \nu }{}^{\lambda }\Sigma
_{AA^{\prime }}^{\nu }-(\Gamma _{\mu A}^{{\footnotesize (T)}}{}^{B}\Sigma
_{BA^{\prime }}^{\lambda }+\text{c.c.}).  \label{63}
\end{equation}%
Therefore, the $\varepsilon $-formalism counterpart of (\ref{61}) must be
spelt out as%
\begin{equation}
\widetilde{\Gamma }_{\mu }+T_{\mu }+\Sigma _{\lambda }^{AA^{\prime
}}\partial _{\mu }\Sigma _{AA^{\prime }}^{\lambda }=0.  \label{64}
\end{equation}

We end this Section by pointing out that the covariant constancy of the $%
\varepsilon $-metric spinors allows the implementation of the $\gamma $%
-formalism statement 
\begin{equation}
\nabla _{\mu }\gamma _{AB}=(\gamma ^{-1}\nabla _{\mu }\gamma )\gamma
_{AB}=(\gamma ^{-1}\widetilde{\nabla }_{\mu }\gamma -\gamma _{\mu }^{%
{\footnotesize (T)}})\gamma _{AB},  \label{65}
\end{equation}%
which yields the expansion 
\begin{equation}
\nabla _{\mu }\gamma _{AB}=(\partial _{\mu }\log \gamma -\gamma _{\mu
})\gamma _{AB}.  \label{66}
\end{equation}%
Since $\nabla _{\mu }\delta _{A}{}^{B}=0$ invariantly, Eqs. (\ref{41}) and (%
\ref{54}) produce the covariant eigenvalue equations%
\begin{equation}
\nabla _{\mu }\gamma _{AB}=i\alpha _{\mu }\gamma _{AB},\text{ }\nabla _{\mu
}\gamma ^{AB}=-i\alpha _{\mu }\gamma ^{AB},  \label{67}
\end{equation}%
along with their complex conjugates and%
\begin{equation}
\alpha _{\mu }=\partial _{\mu }\Phi +2(\Phi _{\mu }+A_{\mu }).  \label{68}
\end{equation}%
It should be noticed that the behaviours specified by Eqs. (\ref{22}) and (%
\ref{52}) guarantee the gauge invariance of $\alpha _{\mu }$. Needless to
say, the occurrence of purely imaginary eigenvalues in (\ref{67}) reflects
the applicability of the $\gamma $-formalism version of the condition (\ref%
{36}).

\section{Spin Curvatures}

The mixed world-spin curvature object associated to either $\vartheta _{\mu
AB}$ occurs in the differential configuration%
\begin{equation}
D_{\mu \nu }\zeta ^{B}=C_{\mu \nu A}{}^{B}\zeta ^{A},  \label{69}
\end{equation}%
where $\zeta ^{A}$ is an arbitrary spin vector and $D_{\mu \nu }$ equals the
operator given by (\ref{2}). Hence, taking the second covariant derivative
of $\zeta ^{A}$ according to one of the expansions (\ref{34}) and performing
some calculational rearrangements, we get the pattern\footnote{%
The object $C_{\mu \nu AB}{}$ carries $48$ real independent components.}%
\begin{equation}
C_{\mu \nu A}{}^{B}=\widetilde{C}_{\mu \nu A}{}^{B}+C_{\mu \nu A}^{%
{\footnotesize (T)}}{}^{B}+\breve{A}_{\mu \nu A}{}^{B}.  \label{70}
\end{equation}%
In Eq. (\ref{70}), the contribution $\widetilde{C}_{\mu \nu A}{}^{B}$ is
identical to the one which occurs in the torsionless framework [16], namely,%
\begin{equation}
\widetilde{C}_{\mu \nu A}{}^{B}=2\partial _{\lbrack \mu }\widetilde{%
\vartheta }_{\nu ]A}{}^{B}-(\widetilde{\vartheta }_{\mu A}{}^{C}\widetilde{%
\vartheta }_{\nu C}{}^{B}-\widetilde{\vartheta }_{\nu A}{}^{C}\widetilde{%
\vartheta }_{\mu C}{}^{B}),  \label{71}
\end{equation}%
and it just arises from%
\begin{equation}
2\widetilde{\nabla }_{[\mu }\widetilde{\nabla }_{\nu ]}\zeta ^{B}=\widetilde{%
C}_{\mu \nu A}{}^{B}\zeta ^{A}.  \label{72}
\end{equation}%
The quantity $C_{\mu \nu A}^{{\footnotesize (T)}}{}^{B}$ takes account of
the torsionfulness of $\vartheta _{\mu AB}$, with its defining expression
being written as%
\begin{equation}
C_{\mu \nu A}^{{\footnotesize (T)}}{}^{B}=2\partial _{\lbrack \mu }\vartheta
_{\nu ]A}^{{\footnotesize (T)}}{}^{B}-(\vartheta _{\mu A}^{{\footnotesize (T)%
}}{}^{C}\vartheta _{\nu C}^{{\footnotesize (T)}}{}^{B}-\vartheta _{\nu A}^{%
{\footnotesize (T)}}{}^{C}\vartheta _{\mu C}^{{\footnotesize (T)}}{}^{B}).
\label{73}
\end{equation}%
In each formalism, the piece $\breve{A}_{\mu \nu A}{}^{B}$ amounts to a
spin-affine entanglement contribution which is typically given by%
\begin{equation}
\breve{A}_{\mu \nu A}{}^{B}=-[(\widetilde{\vartheta }_{\mu A}{}^{C}\vartheta
_{\nu C}^{{\footnotesize (T)}}{}^{B}-\widetilde{\vartheta }_{\nu
A}{}^{C}\vartheta _{\mu C}^{{\footnotesize (T)}}{}^{B})+(\widetilde{%
\vartheta }\vartheta ^{{\footnotesize (T)}}\text{-piece})],  \label{74}
\end{equation}%
where the $\widetilde{\vartheta }\vartheta ^{{\footnotesize (T)}}$-piece
denotes the term that is obtained from the preceding one by interchanging
the roles of the kernel letters $\widetilde{\vartheta }$ and $\vartheta ^{%
{\footnotesize (T)}}$.

Actually, a characteristic curvature splitting for the $\gamma $-formalism
comes about in a straightforward way when we allow $D_{\mu \nu }$ to act
freely upon any Hermitian $\sigma $-object. To see this, we allow for the $%
\gamma $-formalism version of (\ref{63}) and work out the derivative $D_{\mu
\nu }\sigma _{\lambda }^{AA^{\prime }}$. After somewhat lengthy
calculations, we thus obtain the intermediate-stage expansion%
\begin{eqnarray}
D_{\mu \nu }\sigma _{\lambda }^{AA^{\prime }} &=&2\widetilde{\nabla }_{[\mu }%
\widetilde{\nabla }_{\nu ]}\sigma _{\lambda }^{AA^{\prime }}-2(\widetilde{%
\nabla }_{[\mu }T_{\nu ]\lambda }{}^{\rho }+T_{\lambda \lbrack \mu }{}^{\tau
}T_{\nu ]\tau }{}^{\rho })\sigma _{\rho }^{AA^{\prime }}  \notag \\
&&+[2(\widetilde{\nabla }_{[\mu }\gamma _{\nu ]B}^{{\footnotesize (T)}%
}{}^{A}-\gamma _{\lbrack \mu \mid B\mid }^{{\footnotesize (T)}}{}^{C}\gamma
_{\nu ]C}^{{\footnotesize (T)}}{}^{A})\sigma _{\lambda }^{BA^{\prime }}+%
\text{c.c.}].  \label{75}
\end{eqnarray}%
The torsionless second derivative of (\ref{75}) possesses the same form as
that of the traditional framework, that is to say,%
\begin{equation}
2\widetilde{\nabla }_{[\mu }\widetilde{\nabla }_{\nu ]}\sigma _{\lambda
}^{AA^{\prime }}=-\widetilde{R}_{\mu \nu \lambda }{}^{\rho }\sigma _{\rho
}^{AA^{\prime }}+(\widetilde{C}_{\mu \nu B}{}^{A}\sigma _{\lambda
}^{BA^{\prime }}+\text{c.c.}),  \label{76}
\end{equation}%
while the involved world-torsion piece amounts to%
\begin{equation}
2(\widetilde{\nabla }_{[\mu }T_{\nu ]\lambda }{}^{\rho }+T_{\lambda \lbrack
\mu }{}^{\tau }T_{\nu ]\tau }{}^{\rho })=R_{\mu \nu \lambda }^{%
{\footnotesize (T)}}{}^{\rho }+\check{Z}_{\mu \nu \lambda }{}^{\rho },
\label{77}
\end{equation}%
with $\check{Z}_{\mu \nu \lambda }{}^{\rho }$ being the world contribution
(see Eq. (\ref{7}))%
\begin{eqnarray}
\check{Z}_{\mu \nu \lambda }{}^{\rho } &=&-[(\widetilde{\Gamma }_{\mu
\lambda }{}^{\tau }T_{\nu \tau }{}^{\rho }-\widetilde{\Gamma }_{\nu \lambda
}{}^{\tau }T_{\mu \tau }{}^{\rho })+(\widetilde{\Gamma }T\text{-piece})] 
\notag \\
&=&2[(T_{[\mu \mid \tau \mid }{}^{\rho }\widetilde{\Gamma }_{\nu ]\lambda
}{}^{\tau }+(\widetilde{\Gamma }T\text{-piece})],  \label{78}
\end{eqnarray}%
and each of its $\widetilde{\Gamma }T$-pieces coming from an interchange
similar to that of (\ref{74}). In inserting $R_{\mu \nu \lambda
}^{(T)}{}^{\rho }$ into (\ref{77}), it may be convenient to use the trivial
equalities 
\begin{equation}
T_{[\mu \mid \tau \mid }{}^{\rho }T_{\nu ]\lambda }{}^{\tau }=-T_{[\mu \mid
\lambda \mid }{}^{\tau }T_{\nu ]\tau }{}^{\rho }=T_{\lambda \lbrack \mu
}{}^{\tau }T_{\nu ]\tau }{}^{\rho }.  \label{78Lin}
\end{equation}%
Also, Eqs. (\ref{73}) and (\ref{74}) show us that the whole unprimed $\gamma
^{{\footnotesize (T)}}$-contribution of (\ref{75}) reproduces the
corresponding sum $C_{\mu \nu B}^{{\footnotesize (T)}}{}^{A}+\breve{A}_{\mu
\nu B}{}^{A}$ whence, fitting pieces together, yields the expression%
\begin{eqnarray}
D_{\mu \nu }\sigma _{\lambda }^{AA^{\prime }} &=&-(\widetilde{R}_{\mu \nu
\lambda }{}^{\rho }+R_{\mu \nu \lambda }^{{\footnotesize (T)}}{}^{\rho }+%
\check{Z}_{\mu \nu \lambda }{}^{\rho })\sigma _{\rho }^{AA^{\prime }}  \notag
\\
&&+[(\widetilde{C}_{\mu \nu B}{}^{A}+C_{\mu \nu B}^{{\footnotesize (T)}%
}{}^{A}+\breve{A}_{\mu \nu B}{}^{A})\sigma _{\lambda }^{BA^{\prime }}+\text{%
c.c.}],  \label{79}
\end{eqnarray}%
which suggests defining the formal expansion%
\begin{equation}
D_{\mu \nu }\sigma _{\lambda }^{AA^{\prime }}=-R_{\mu \nu \lambda }{}^{\rho
}\sigma _{\rho }^{AA^{\prime }}+(C_{\mu \nu B}{}^{A}\sigma _{\lambda
}^{BA^{\prime }}+\text{c.c.}),  \label{80}
\end{equation}%
in agreement with Eqs. (\ref{7}) and (\ref{70}). Then, transvecting (\ref{80}%
) with $\sigma _{CA^{\prime }}^{\lambda }$ leads to%
\begin{equation}
2C_{\mu \nu A}{}^{B}+\delta _{A}{}^{B}C_{\mu \nu A^{\prime }}{}^{A^{\prime
}}-\sigma _{AA^{\prime }}^{\lambda }\sigma ^{\rho BA^{\prime }}R_{\mu \nu
\lambda \rho }{}=0,  \label{81}
\end{equation}%
which, in turn, brings about the property%
\begin{equation}
\func{Re}C_{\mu \nu A}{}^{A}=0,  \label{82}
\end{equation}%
provided that $R_{\mu \nu \lambda }{}^{\lambda }\equiv 0$. Consequently,
since the contracted quadratic pieces of both (\ref{71}) and (\ref{73})\
vanish identically together with $\breve{A}_{\mu \nu B}{}^{B}$, we get the
additivity relation%
\begin{equation}
C_{\mu \nu A}{}^{A}=\widetilde{C}_{\mu \nu A}{}^{A}+C_{\mu \nu A}^{%
{\footnotesize (T)}}{}^{A},  \label{83}
\end{equation}%
along with the purely imaginary twelve-parameter contribution%
\begin{equation}
C_{\mu \nu A}{}^{A}=-2iF_{\mu \nu }\doteqdot -2i(\widetilde{F}_{\mu \nu
}+F_{\mu \nu }^{{\footnotesize (T)}}),  \label{84}
\end{equation}%
where%
\begin{equation}
\widetilde{F}_{\mu \nu }\doteqdot 2\partial _{\lbrack \mu }\Phi _{\nu ]},%
\text{ }F_{\mu \nu }^{{\footnotesize (T)}}\doteqdot 2\partial _{\lbrack \mu
}A_{\nu ]},  \label{85}
\end{equation}%
with Eqs. (\ref{40})-(\ref{42}) having been employed for expressing (\ref{85}%
).

It is of interest to recast the pieces of Eq. (\ref{85}) as%
\begin{equation}
\widetilde{F}_{\mu \nu }\doteqdot 2\widetilde{\nabla }_{[\mu }\Phi _{\nu ]},%
\text{ }F_{\mu \nu }^{{\footnotesize (T)}}\doteqdot 2(\nabla _{\lbrack \mu
}A_{\nu ]}+T_{\mu \nu }{}^{\lambda }A_{\lambda }).  \label{86}
\end{equation}%
Hence, lowering the index $B$ of Eq. (\ref{81}), gives the splitting%
\begin{equation}
C_{\mu \nu AB}{}=\frac{1}{2}\sigma _{AA^{\prime }}^{\lambda }{}\sigma
_{B}^{\rho A^{\prime }}R_{\mu \nu \lambda \rho }{}-iF_{\mu \nu }\gamma _{AB},
\label{87}
\end{equation}%
which recovers in the $\gamma $-formalism the number of real independent
components of $C_{\mu \nu AB}$ as $36+12$. Because of the relation (\ref{26}%
), the $R$-configuration of (\ref{87}) bears symmetry in $A$ and $B$ such
that 
\begin{equation}
C_{\mu \nu (AB)}{}=\frac{1}{2}\sigma _{A^{\prime }A}^{\lambda }{}\sigma
_{B}^{\rho A^{\prime }}R_{\mu \nu \lambda \rho }{}.  \label{87Lin}
\end{equation}

The derivation of the $\varepsilon $-formalism counterpart of Eq. (\ref{87})
is carried out along the same lines as those yielding (\ref{80}), but now we
have to require%
\begin{equation}
\nabla _{\lbrack \mu }(\Upsilon _{\nu ]}\Sigma _{\lambda }^{AA^{\prime
}})=0\Leftrightarrow \partial _{\lbrack \mu }\Upsilon _{\nu ]}=T_{\mu \nu
}{}^{\lambda }\Upsilon _{\lambda }.  \label{88}
\end{equation}%
In the classical framework, a similar requirement is also made which neatly
fits in with the transformation law for the pertinent $\Upsilon _{\mu }$.
Here, we naively ascribe a gauge-invariant character to (\ref{88}) by
choosing gauge matrices that possess constant modulus determinants, in which
case we may write down the $\varepsilon $-expression%
\begin{equation}
C_{\mu \nu AB}{}=\frac{1}{2}\Sigma _{AA^{\prime }}^{\lambda }{}\Sigma
_{B}^{\rho A^{\prime }}R_{\mu \nu \lambda \rho }{}-iF_{\mu \nu }\varepsilon
_{AB}.  \label{89}
\end{equation}%
In the $\gamma $-formalism, we thus have the tensor law 
\begin{equation}
C_{\mu \nu AB}^{\prime }={}\Lambda _{A}{}^{C}\Lambda _{B}{}^{D}C_{\mu \nu
CD}=\Delta _{{\small \Lambda }}C_{\mu \nu AB},  \label{90}
\end{equation}%
whereas the object $C_{\mu \nu AB}{}$ for the $\varepsilon $-formalism must
be taken as an invariant spin-tensor density of weight $-1$, whence we also
have 
\begin{equation}
C_{\mu \nu AB}^{\prime }{}=(\Delta _{{\small \Lambda }})^{-1}\Lambda
_{A}{}^{C}\Lambda _{B}{}^{D}C_{\mu \nu CD}=C_{\mu \nu AB}.  \label{91}
\end{equation}

The curvature spinors for either formalism enter the bivector decomposition
of the respective $C_{\mu \nu AB}$. We have, in effect,%
\begin{equation}
S_{AA^{\prime }}^{\mu }S_{BB^{\prime }}^{\nu }C_{\mu \nu CD}=M_{A^{\prime
}B^{\prime }}\varpi _{ABCD}+M_{AB}\varpi _{A^{\prime }B^{\prime }CD},
\label{92}
\end{equation}%
along with the definitions%
\begin{equation}
\varpi _{ABCD}=\varpi _{(AB)CD}\doteqdot \frac{1}{2}S_{A^{\prime }A}^{\mu
}S_{B}^{\nu A^{\prime }}C_{\mu \nu CD}  \label{93}
\end{equation}%
and%
\begin{equation}
\varpi _{A^{\prime }B^{\prime }CD}=\varpi _{(A^{\prime }B^{\prime
})CD}\doteqdot \frac{1}{2}S_{AA^{\prime }}^{\mu }S_{B^{\prime }}^{\nu
A}C_{\mu \nu CD},  \label{94}
\end{equation}%
with the symmetries shown by (\ref{93}) and (\ref{94}) being once again
ensured by Eq. (\ref{26}). Each of the $\varpi $-curvatures of (\ref{92})
obviously contributes $24$ real independent components to $C_{\mu \nu CD}$.
Owing to the gauge behaviours of the metric spinors and $C$-objects, the
curvature spinors for the $\gamma $-formalism are subject to the tensor laws%
\begin{equation}
\varpi _{ABCD}^{\prime }=\Lambda _{A}{}^{L}\Lambda _{B}{}^{M}\Lambda
_{C}{}^{R}\Lambda _{D}{}^{S}\varpi _{LMRS}=(\Delta _{{\small \Lambda }%
})^{2}\varpi _{ABCD}  \label{95}
\end{equation}%
and 
\begin{equation}
\varpi _{A^{\prime }B^{\prime }CD}^{\prime }=\Lambda _{A^{\prime
}}{}^{L^{\prime }}\Lambda _{B^{\prime }}{}^{M^{\prime }}\Lambda
_{C}{}^{R}\Lambda _{D}{}^{S}\varpi _{L^{\prime }M^{\prime }RS}=\hspace{0.01cm%
}\rho ^{2}\varpi _{A^{\prime }B^{\prime }CD},  \label{96}
\end{equation}%
while the ones for the $\varepsilon $-formalism are invariant spin-tensor
densities prescribed by 
\begin{equation}
\varpi _{ABCD}^{\prime }=(\Delta _{{\small \Lambda }})^{-2}\Lambda
_{A}{}^{L}\Lambda _{B}{}^{M}\Lambda _{C}{}^{R}\Lambda _{D}{}^{S}\varpi
_{LMRS}=\varpi _{ABCD}  \label{97}
\end{equation}%
and 
\begin{equation}
\varpi _{A^{\prime }B^{\prime }CD}^{\prime }=\hspace{0.05cm}\rho
^{-2}\Lambda _{A^{\prime }}{}^{L^{\prime }}\Lambda _{B^{\prime
}}{}^{M^{\prime }}\Lambda _{C}{}^{R}\Lambda _{D}{}^{S}\varpi _{L^{\prime
}M^{\prime }RS}=\varpi _{A^{\prime }B^{\prime }CD}.  \label{98}
\end{equation}

A glance at Eqs. (\ref{83}), (\ref{84}) and (\ref{89}) tells us that the
contracted curvature spinors $(\varpi _{ABC}{}^{C},\varpi _{A^{\prime
}B^{\prime }C}{}^{C})$ for both formalisms constitute the bivector
decomposition%
\begin{equation}
-2iS_{AA^{\prime }}^{\mu }S_{BB^{\prime }}^{\nu }F_{\mu \nu }=M_{A^{\prime
}B^{\prime }}\varpi _{ABC}{}^{C}+M_{AB}\varpi _{A^{\prime }B^{\prime
}C}{}^{C},  \label{99}
\end{equation}%
in addition to satisfying the property%
\begin{equation}
\varpi _{ABC}{}^{C}=\widetilde{\varpi }_{ABC}{}^{C}+\varpi _{ABC}^{%
{\footnotesize (T)}}{}^{C},\text{ }\varpi _{A^{\prime }B^{\prime }C}{}^{C}=%
\widetilde{\varpi }_{A^{\prime }B^{\prime }C}{}^{C}+\varpi _{A^{\prime
}B^{\prime }C}^{{\footnotesize (T)}}{}^{C}.  \label{100}
\end{equation}%
Each of the pairs $(\widetilde{\varpi }_{ABC}{}^{C},\widetilde{\varpi }%
_{A^{\prime }B^{\prime }C}{}^{C})$ and $(\varpi _{ABC}^{{\footnotesize (T)}%
}{}^{C},\varpi _{A^{\prime }B^{\prime }C}^{{\footnotesize (T)}}{}^{C})$ is
now taken to contribute $6$ real independent components to the overall $%
C_{\mu \nu A}{}^{A}$ of either formalism. Hence, making use of the torsional
expression of (\ref{86}) along with the prescriptions%
\begin{equation}
T_{AA^{\prime }BB^{\prime }}{}^{CC^{\prime }}A_{CC^{\prime }}=M_{A^{\prime
}B^{\prime }}\tau _{AB}{}^{CC^{\prime }}A_{CC^{\prime }}+\text{c.c.}
\label{101}
\end{equation}%
and%
\begin{equation}
\tau _{AB}{}^{CC^{\prime }}\doteqdot \frac{1}{2}T_{(AB)D^{\prime
}}{}^{D^{\prime }CC^{\prime }},  \label{102}
\end{equation}%
we obtain the relationships%
\begin{equation}
\varpi _{ABC}^{{\footnotesize (T)}}{}^{C}=2i(\nabla _{(A}^{C^{\prime
}}A_{B)C^{\prime }}-2\tau _{AB}{}^{CC^{\prime }}A_{CC^{\prime }})
\label{103}
\end{equation}%
and\footnote{%
We emphasize that, within our framework, $\widetilde{\nabla }_{\mu
}S_{\lambda }^{AA^{\prime }}\neq 0$.}%
\begin{equation}
\varpi _{A^{\prime }B^{\prime }C}^{{\footnotesize (T)}}{}^{C}=2i(\nabla
_{(A^{\prime }}^{C}A_{B^{\prime })C}-2\tau _{A^{\prime }B^{\prime
}}{}^{CC^{\prime }}A_{CC^{\prime }}),  \label{104}
\end{equation}%
together with, say,%
\begin{equation}
\widetilde{\varpi }_{ABC}{}^{C}=2i(\nabla _{(A}^{C^{\prime }}\Phi
_{B)C^{\prime }}-2\tau _{AB}{}^{CC^{\prime }}\Phi _{CC^{\prime }}).
\label{105}
\end{equation}%
Either of $\tau _{AB}{}^{CC^{\prime }}$ and $\tau _{A^{\prime }B^{\prime
}}{}^{CC^{\prime }}$ recovers the number of independent components of $%
T_{\mu \nu }{}^{\lambda }$ as $4\times 6$. We must point out that the
contracted curvature spinors for both formalisms obey the simultaneous
conjugation relations%
\begin{equation}
\varpi _{ABC}{}^{C}=-\hspace{1pt}\varpi _{ABC^{\prime }}{}^{C^{\prime }},%
\text{ }\varpi _{A^{\prime }B^{\prime }C}{}^{C}=-\hspace{1pt}\varpi
_{A^{\prime }B^{\prime }C^{\prime }}{}^{C^{\prime }}.  \label{106}
\end{equation}

The Riemann curvature structure of $\mathfrak{M}$ as defined by Eqs. (\ref{6}%
)-(\ref{8}) can be completely reinstated from the symmetric pair 
\begin{equation}
\mathbf{R}=(\varpi _{AB(CD)},\text{ }\varpi _{A^{\prime }B^{\prime }(CD)}),
\label{107}
\end{equation}%
with each entry of which thus carrying $18$ real independent components. The
torsionless version of this statement was established in Ref. [18] out of
utilizing some elementary metric formulae that may be formally applied to
the case of (\ref{107}) too. We can therefore recover as $18+6$ the number
of degrees of freedom of each of the $\varpi $-spinors carried by Eq. (\ref%
{92}). In both formalisms, we then have the gauge-covariant expression 
\begin{equation}
R_{AA^{\prime }BB^{\prime }CC^{\prime }DD^{\prime }}=\hspace{-1pt}%
(M_{A^{\prime }B^{\prime }}M_{C^{\prime }D^{\prime }}\varpi _{AB(CD)}\hspace{%
-1pt}+M_{AB}M_{C^{\prime }D^{\prime }}\varpi _{A^{\prime }B^{\prime }(CD)})+%
\text{c.c.},  \label{108}
\end{equation}%
such that%
\begin{equation}
\varpi _{AB(CD)}=\frac{1}{4}M^{A^{\prime }B^{\prime }}M^{C^{\prime
}D^{\prime }}R_{AA^{\prime }BB^{\prime }CC^{\prime }DD^{\prime }}
\label{109}
\end{equation}%
and%
\begin{equation}
\varpi _{A^{\prime }B^{\prime }(CD)}=\frac{1}{4}M^{AB}M^{C^{\prime
}D^{\prime }}R_{AA^{\prime }BB^{\prime }CC^{\prime }DD^{\prime }},
\label{110}
\end{equation}%
with the number of independent components of $R_{\mu \nu \lambda \sigma }$
accordingly appearing as $18+18$. It should be evident that the symmetries
brought out by the configurations (\ref{93}), (\ref{94}) and (\ref{108})
just correspond to the skew symmetry in the indices of the pairs $\mu \nu $
and $\lambda \sigma $ borne by $R_{\mu \nu \lambda \sigma }$. These are
indeed the only symmetries carried by the curvature spinors (\ref{107}).
With the help of Eq. (\ref{30}), we write the first-left dual of (\ref{108})
as%
\begin{equation}
^{\ast }R_{AA^{\prime }BB^{\prime }CC^{\prime }DD^{\prime
}}=[(-i)(M_{A^{\prime }B^{\prime }}M_{C^{\prime }D^{\prime }}\varpi _{AB(CD)}%
\hspace{-1pt}-M_{AB}M_{C^{\prime }D^{\prime }}\varpi _{A^{\prime }B^{\prime
}(CD)})]+\text{c.c.},  \label{111}
\end{equation}%
whence the pair (\ref{107}) may be directly obtained from the affine
correlations of Section 3.

In either formalism, the number of degrees of freedom of $\varpi _{AB(CD)}$
becomes transparently visible when we put into effect the definitions%
\begin{equation}
\varpi _{AB(CD)}\doteqdot \text{X}_{ABCD},\text{ }\varpi _{A^{\prime
}B^{\prime }(CD)}\doteqdot \Xi _{A^{\prime }B^{\prime }CD},  \label{112}
\end{equation}%
along with the reduction device\footnote{%
In either of the old formalisms, the X-spinor carries $18-6-1$ degrees of
freedom while the $\Xi $-spinor carries $10-1$ and bears Hermiticity. Within
our framework, either $\Xi $-spinor can not be associated to any world
tensor.}%
\begin{align}
\hspace{-0.1cm}\hspace{-0.01cm}\text{X}_{ABCD}\hspace{-0.07cm}& =\hspace{%
-0.07cm}\text{X}_{(ABCD)}-\frac{1}{4}(M_{AB}\text{X}^{L}{}_{(LCD)}+M_{AC}%
\text{X}^{L}{}_{(LBD)}+M_{AD}\text{X}^{L}{}_{(LBC)})  \notag \\
& \hspace{-0.07cm}-\frac{1}{3}(M_{BC}\text{X}^{L}{}_{A(LD)}+M_{BD}\text{X}%
^{L}{}_{A(LC)})-\frac{1}{2}M_{CD}\text{X}_{AB}{}^{L}{}_{L}.  \label{113}
\end{align}%
Some calculations then yield the explicit expansion%
\begin{equation}
\text{X}_{ABCD}\hspace{-0.07cm}=\hspace{-0.07cm}\Psi _{ABCD}-M_{(A\mid
(C}\xi _{D)\mid B)}-\frac{1}{3}\varkappa M_{A(C}M_{D)B},  \label{114}
\end{equation}%
together with the individual pieces%
\begin{equation}
\hspace{-0.07cm}\Psi _{ABCD}=\text{X}_{(ABCD)}\hspace{-0.07cm},\text{ }\xi
_{AB}=\text{X}^{M}{}_{(AB)M},\text{ }\varkappa =\text{X}_{LM}{}^{LM}.
\label{115}
\end{equation}%
It is clear that the world-covariant character of $C_{\mu \nu AB}{}$ and the
behaviours described by Eqs. (\ref{95})-(\ref{98}), assure that $\varkappa $
is a world-spin invariant in both formalisms. Moreover, since $^{\ast
}R^{\lambda }{}_{\mu \lambda \nu }\neq 0$, we must regard $\varkappa $ as a
complex quantity. In essence, the factor $1/3$ that occurs in the reduction
of X$_{ABCD}$ we have deduced, is due to the lack of symmetry of the piece X$%
^{M}{}_{ABM}$, in contraposition to the torsionless framework wherein the
counterpart of the $\varkappa $-term of (\ref{114}) carries a factor $2/3$
because $\xi _{AB}\equiv 0$ thereabout. Hence, the pieces of (\ref{115})
contribute $(5,3,1)$ complex independent components to $\varpi _{AB(CD)}$,
respectively. This component prescription was exhibited for the first time
in Ref. [31]. For the Ricci tensor and scalar of $\nabla _{\mu }$, we thus
get the expressions%
\begin{equation}
R_{AA^{\prime }BB^{\prime }}=M_{AB}M_{A^{\prime }B^{\prime }}\func{Re}%
\varkappa -[(M_{A^{\prime }B^{\prime }}\xi _{AB}+\Xi _{A^{\prime }B^{\prime
}AB})+\text{c.c.}]  \label{116}
\end{equation}%
and%
\begin{equation}
R=4\func{Re}\varkappa ,  \label{117}
\end{equation}%
with Eq. (\ref{116}) recovering the number of degrees of freedom of $R_{\mu
\nu }$ as $1+6+9$. Likewise, for the contracted first-left dual $^{\ast
}R^{\lambda }{}_{\mu \lambda \nu }$, we have%
\begin{equation}
^{\ast }R^{CC^{\prime }}{}_{AA^{\prime }CC^{\prime }BB^{\prime
}}=[i(M_{A^{\prime }B^{\prime }}\xi _{AB}-\frac{1}{2}M_{AB}M_{A^{\prime
}B^{\prime }}\varkappa -\Xi _{A^{\prime }B^{\prime }AB})]+\text{c.c.},
\label{118}
\end{equation}%
whence%
\begin{equation}
^{\ast }R{}_{\mu \nu }{}^{\mu \nu }=4\func{Im}\varkappa .  \label{119}
\end{equation}

We shall now proceed to deriving the spinor version of Eqs. (\ref{11})\ and (%
\ref{12}). Let us write the dual spinor torsion%
\begin{equation}
^{\ast }T_{AA^{\prime }BB^{\prime }CC^{\prime }}=i(M_{AB}\tau _{A^{\prime
}B^{\prime }CC^{\prime }}-\text{c.c.}),  \label{120}
\end{equation}%
with its $\tau $-pieces being defined explicitly by Eq. (\ref{102}). In the $%
\gamma $-formalism, the world derivative $\nabla ^{\lambda }{}^{\ast
}T_{\lambda \mu \nu }$ then corresponds to%
\begin{equation}
\nabla ^{AA^{\prime }\text{ }\ast }T_{AA^{\prime }BB^{\prime }CC^{\prime
}}=i[(\nabla _{B}^{A^{\prime }}\tau _{A^{\prime }B^{\prime }CC^{\prime
}}+i\alpha _{B}^{A^{\prime }}\tau _{A^{\prime }B^{\prime }CC^{\prime }})-%
\text{c.c.}],  \label{121}
\end{equation}%
where Eq. (\ref{67}) has been utilized. The $\varepsilon $-formalism version
of (\ref{121}) may be obtained by simply dropping the $\alpha $-term from
it, namely,%
\begin{equation}
\nabla ^{AA^{\prime }\text{ }\ast }T_{AA^{\prime }BB^{\prime }CC^{\prime
}}=i(\nabla _{B}^{A^{\prime }}\tau _{A^{\prime }B^{\prime }CC^{\prime }}-%
\text{c.c.}).  \label{122}
\end{equation}%
We next write the $^{\ast }TT$-kernel of Eq. (\ref{11}) as%
\begin{eqnarray}
&&^{\ast }T_{BB^{\prime }}{}^{DD^{\prime }MM^{\prime }}T_{DD^{\prime
}MM^{\prime }CC^{\prime }}  \notag \\
&=&i[(\tau _{B}{}^{DM}{}_{B^{\prime }}\tau _{DMCC^{\prime }}-\text{c.c.}%
)-(\tau _{BD}{}^{DD^{\prime }}\tau _{B^{\prime }D^{\prime }CC^{\prime }}-%
\text{c.c.})].  \label{123}
\end{eqnarray}%
In both formalisms, the combination of Eqs. (\ref{121})-(\ref{123}) with the
expression (\ref{118}) enables one to express the pair (\ref{107}) in terms
of spin-torsion constituents and their derivatives, as had been alluded to
in Ref. [31]. For the contracted derivative involved in Eq. (\ref{12}), we
have the reduced contribution%
\begin{eqnarray}
&&M^{C^{\prime }D^{\prime }}\nabla ^{AA^{\prime }}{}^{\ast }R_{AA^{\prime
}BB^{\prime }CC^{\prime }DD^{\prime }}  \notag \\
&=&(-2i)(\nabla _{B^{\prime }}^{A}\text{X}_{ABCD}-2i\alpha _{B^{\prime }}^{A}%
\text{X}_{ABCD}-\nabla _{B}^{A^{\prime }}\Xi _{A^{\prime }B^{\prime }CD}),
\label{124}
\end{eqnarray}%
together with the complex conjugate of it. Whence, the Bianchi identity can
be recovered by combining (\ref{124}) with%
\begin{eqnarray}
&&M^{C^{\prime }D^{\prime }\text{ }\ast }T_{BB^{\prime }}{}^{LL^{\prime
}MM^{\prime }}R_{LL^{\prime }MM^{\prime }CC^{\prime }DD^{\prime }}  \notag \\
&=&2i[(\tau _{B^{\prime }L^{\prime }}{}^{LL^{\prime }}\text{X}_{LBCD}-\tau
_{B^{\prime }}{}^{(L^{\prime }M^{\prime })}{}_{B}\Xi _{L^{\prime }M^{\prime
}CD})  \notag \\
&&+(\tau _{B}{}^{(LM)}{}_{B^{\prime }}\text{X}_{LMCD}-\tau
_{BL}{}^{LL^{\prime }}\Xi _{L^{\prime }B^{\prime }CD})].  \label{125}
\end{eqnarray}

\section{Concluding Remarks and Outlook}

According to the classical world geometry, any torsion tensors may be
suppressed from general affine prescriptions while symmetric affinities may
not. This is because the inhomogeneous parts coming from partial derivatives
are strictly cancelled by those carried by symmetric connexions. Thus, the
geometric adequacy of the equality (\ref{7}) stems from the world-tensor
character of Eq. (\ref{77}). It is worth remarking that any non-contracted
curvatures may not enjoy the additivity property. This feature has
necessarily to be carried over to $R_{\mu \nu }$ and $R$ as the
configuration (\ref{78}) appropriately yields both $\check{Z}_{\lambda \mu
}{}^{\lambda }{}_{\nu }\neq 0$ and $\check{Z}_{\mu \nu }{}^{\mu \nu }\neq 0$%
. Therefore, we can roughly say that the property (\ref{83}) ceases holding
for the world case.

Since any two-component spinor approach to curved spacetimes should formally
resemble the traditional world ones, we can infer that any torsional
affinities like those we have defined in Section 3 must always be
accompanied by suitable torsionless $\gamma \varepsilon $-affine
contributions. It is upon this fact that the genuineness of the gauge laws (%
\ref{47}) and (\ref{48}) rests. Hence, as the contracted affinity $\vartheta
_{\mu A}^{{\footnotesize (T)}}{}^{A}$ for either formalism has been chosen
such that $\func{Re}\vartheta _{\mu }^{{\footnotesize (T)}}=0$, we can
conclude that the limiting procedure which could be implemented hereabout
takes world and spin torsion contributions to vanish independently of one
another.

We realize that the Bianchi identity as exhibited by Eqs. (\ref{124}) and (%
\ref{125}) could not only bring out the most characteristic form of
two-component spinor couplings between curvatures and torsion, but could
also afford the field equations which control the propagation of gravitons
in torsional environments. With regard to this latter situation, the
torsionful extension of the differential calculational techniques used in
Refs. [22, 23] for deriving the wave equations of the torsionless framework,
would be of the utmost significance. Such techniques may of course be
additionally utilized to describe the interaction in the presence of torsion
between the cosmic microwave background and Dirac particles. In our view,
the inner structure of the torsional spinor formalisms just constructed
could provide locally a realistic description of the cosmic dark energy
through a gauge-invariant potential like that defined by Eq. (\ref{37}),
while at the same time assigning geometrically a clear physical meaning to
the right-hand side of Einstein-Cartan's field equations. In spacetimes
having $R=0$, the expressions (\ref{116}) and (\ref{117}) for either
formalism particularly yield a purely imaginary $\varkappa $-quantity
together with a traceless energy-momentum tensor $E_{\mu \nu }$ and the
statement%
\begin{equation*}
(M_{A^{\prime }B^{\prime }}\xi _{AB}+\Xi _{A^{\prime }B^{\prime }AB})+\text{%
c.c.}=\kappa E_{AA^{\prime }BB^{\prime }}.
\end{equation*}

\end{document}